# A possible spin Jahn-Teller material: ordered pseudobrookite FeTi$_2$O$_5$

Hao-Hang Xu[1], Jian Liu[4], L.L. Tao[1], Xian-Jie Wang [1,*], Sergey V. Streltsov[2,3*] and Yu Sui[1,4,*]

[1]*School of Physics, Harbin Institute of Technology, Harbin 150001, China*

[2]*Institute of Metal Physics, S. Kovalevskoy Street 18, Ekaterinburg 620990, Russia*

[3]*Ural Federal University, Mira Street 19, Ekaterinburg 620002, Russia*

[4]*Laboratory for Space Environment and Physical Sciences, Harbin Institute of Technology, Harbin 150001, China*

## Abstract

We investigated the spin-lattice coupling in orthorhombic pseudobrookite FeTi$_2$O$_5$ single crystal with highly ordered Fe$^{2+}$/Ti$^{4+}$ occupation, which consists of quasi-1D $S=2$ chains running along *a*-axis. Both the magnetization and specific heat measurements confirm that the antiferromagnetic phase transition of FeTi$_2$O$_5$ occurs at $T_N$ = 42 K. The structural distortions were also observed around $T_N$ in the thermal expansion $\Delta L/L(T)$ data. Moreover, the magnetic field was found to strongly affect the thermal expansion both along chains and in the perpendicular direction clearly signaling a substantial magnetoelastic coupling, which was recently proposed to be the origin of a rare spin Jahn-Teller effect, when frustration is lifted via additional lattice distortions. Experimentally observed change in the thermal conductivity slope around $T_N$ is usually associated with the orbital ordering, but DFT+U calculations do not detect modification of the orbital structure across the transition. However, the first-principles calculation results confirm that FeTi$_2$O$_5$ is a quasi-1D magnet with a ratio of frustrating inter-chain to intra-chain exchanges $J'/J$ = 0.03 and a substantial single-ion anisotropy (A = 4K) of easy-axis type making this material interesting for studying quantum criticality in transverse magnetic fields.

## I. INTRODUCTION

The Jahn-Teller (JT) effect is a spontaneous structural distortion that lifts the orbital degeneracy of the *d*-electron, resulting in the orbital ordering and causing various fascinating phenomena, e.g. it can affect magnetic structure [1], lead to the orbital-selective Mott transition [2] or the reduction of effective dimensionality [3]. In analogy to the JT effect, Yamashita *et al.* proposed the spin JT effect, when the spin degeneracy in geometrical frustration spin systems is released by the structural distortion through the spin-lattice coupling [4]. While this mechanism was initially proposed for $ZnV_2O_4$ and $MgV_2O_4$, it was later also applied to other spinels [5–7].

The orthorhombic pseudobrookite $CoTi_2O_5$ was recently proposed by Kirschner *et al.* as another candidate for spin JT effect [8]. It was suggested that the triangular geometrical frustration exists in the orthorhombic pseudobrookite $CoTi_2O_5$ and an unexpected long-range antiferromagnetic ordering of $Co^{2+}$ at $T_N \sim 26$ K is due to an additional structural distortion induced by the spin-lattice coupling through the spin JT effect [8]. Combined the X-ray diffraction measurements on polycrystalline samples with the density functional theory calculations, Lang *et al.* demonstrated that the pseudobrookite $FeTi_2O_5$ has the same crystal and magnetic structures as $CoTi_2O_5$, so the triangular geometrical frustration also exists in $FeTi_2O_5$, as shown in Fig.1(a), and a structural distortion is also expected in $FeTi_2O_5$ [9], but no direct experimental evidence was found. This is not surprising since distortions related to lowering of the crystal structure due to the spin JT effect are expected to be extremely small and are unlikely to be measured by neutron or X-ray diffraction techniques. This remains the situation in multiferroics, where electric polarization is measurable, but corresponding distortions often remain experimentally elusive. In addition, compounds with octahedrally coordinated $Fe^{2+}$ could also exhibit various interesting physical properties due to the orbital ordering of $Fe^{2+}$. For instance, the orbital ordering of $Fe^{2+}$ accompanied by a structural distortion in perovskite $KFeF_3$ opens a band gap in the electronic band structure, leading to a metal-insulator transition [10]. Moreover, there is an interplay between charge, spin, and lattice degrees of freedom in famous $Fe_3O_4$, which results in formation of trimerons and Verwey transition [11]. Therefore, it is interesting to study coupling between all these degrees of freedom in $FeTi_2O_5$ as well.

There are two main structural units in the pseudobrookite $FeTi_2O_5$: $Fe^{2+}O_6$ and $Ti^{4+}O_6$ octahedrons. Like in many other pseudobrookites [12,13], the cations order-disorder also occurs in $FeTi_2O_5$ (*Cmcm* space group) due to the existence of the inequivalent $M1O_6$ and $M2O_6$ octahedrons, where $M$ represents the $Fe^{2+}/Ti^{4+}$ ions. For the highly ordered $FeTi_2O_5$, $Fe^{2+}(r=0.76$ Å) and $Ti^{4+}(r=0.68$ Å) will occupy the larger $M1O_6$ octahedron and the smaller $M2O_6$ octahedron, respectively. However, since the 3*d* shell of $Ti^{4+}$ ions is completely unoccupied, the unusual magnetic and structural properties of this material are believed to be related to Fe $3d$ electrons [9]. Therefore, there will be no magnetic frustration in $FeTi_2O_5$ once the vertices of the triangle in Fig.1 (a) are occupied by nonmagnetic $Ti^{4+}$. Since the frustration is necessary for the spin JT effect, highly ordered $Fe^{2+}/Ti^{4+}$ occupancy is needed for investigating the spin JT effect in $FeTi_2O_5$. On the other hand, for the completely ordered $FeTi_2O_5$, the distance between adjacent $Fe^{2+}$ along the $b$ and $c$ directions is much larger than that along the $a$-axis ($a$ = 3.74098 Å, $b$ = 9.7609 Å, $c$ = 10.0914 Å) [9], so the 1D $S$ = 2 chains of $Fe^{2+}$ are formed along the $a$-axis. The appearance of the 1D spin chain can result in various fascinating phenomena, e.g. the Bose-Einstein condensation, formation of quantum spin-liquid states, and Haldane phase [14-16], making $FeTi_2O_5$ attractive for studying also from this perspective.

In this paper, using the floating zone method, the $FeTi_2O_5$ single crystal with highly ordered $Fe^{2+}/Ti^{4+}$ occupation was successfully grown. Through thermal expansion and magnetostriction measurements, we found that the magnetoelastic coupling exists in $FeTi_2O_5$.

## II. EXPERIMENTAL AND CALCULATION DETAILS

A $FeTi_2O_5$ single crystal was grown by the optical floating zone technique in an image furnace with two ellipsoidal mirrors (IR Image Furnace G3, Quantum Design Japan). The crystal was grown in the pure Ar atmosphere with a growth rate of 1 mm/h. $MgTi_2O_5$ single crystal was also prepared under the same condition in order to estimate the lattice specific heat of $FeTi_2O_5$. The phase purity was identified by the X-ray diffraction (XRD, Aeris, Cu$K\alpha_1$ radiation) and the X-ray Laue back diffraction was used to confirm the quality of the crystal and determine the crystal principal axes. The magnetic susceptibility and magnetization measurements were performed by using a commercial superconducting quantum interference device magnetometer (MPMS3). The measurement of specific heat $C(T)$ was performed by the physical property measurement system (PPMS, DynaCool-14 T). Both the thermal expansion $\Delta L/L_0(T)$ and the magnetostriction $\Delta L/L_0(H)$ were performed in PPMS by using AH 2550A capacitance dilatometer that was calibrated with 99.999% pure Cu and Al rods. The thermal conductivity $\kappa(T)$ is measured in PPMS with the "one heater, two thermometers" method.

We used density functional theory calculations as realized in the VASP package [17] with generalized gradient approximation (GGA) and the exchange-correlation potential in PBE form [18] to study the electronic and magnetic properties of $FeTi_2O_5$. The correlation effects were taken into account by the GGA+U approach [19] with *U-$J_H$= 4* eV for Fe and *U-$J_H$= 2.5* eV, close to typical values of Hubbard repulsion (*U*) and intra-atomic exchange used in the literature [20,21]. All calculations were performed for the

9x9x5 mesh of the Brillouin zone, the convergence criterion was set to $10^{-5}$ eV, while the cut-off energy was chosen to be 500 eV. Following Wigner-Seitz radii were chosen for calculations of atomic charges and magnetic moments: $R_{Fe}$=1.302 Å, $R_{Ti}$=1.323 Å, and $R_O$=0.82 Å.

# III. RESULTS AND DISCUSSION

## A. X-ray diffraction and magnetic susceptibility

To identify the degree of the ordering of $Fe^{2+}/Ti^{4+}$ in $FeTi_2O_5$, the powder XRD was measured and the Rietveld refinement was performed to identify the degree of the ordering of $Fe^{2+}/Ti^{4+}$ in $FeTi_2O_5$, as shown in Fig.1(b). The refined lattice parameters for room temperature are $a$ = 3.74014(2) Å, $b$ = 9.76375(6) Å, and $c$ = 10.08496(7) Å, as listed in Table I, consistent with the results reported before [9]. From the refinement results, we can also get the information of the $Fe^{2+}/Ti^{4+}$ disorder parameter $X$ in $FeTi_2O_5$, which is defined as the atomic concentration of $Ti^{4+}$ in TM1 sites [22]. For the grown $FeTi_2O_5$ single crystal, $X$ is only 0.12, which is smaller than 0.14 for the well-studied pseudo-brookite $MgTi_2O_5$ ceramics with highly ordered $Mg^{2+}/Ti^{4+}$ occupation [22]. So, in our $FeTi_2O_5$ single crystal, most of the $Fe^{2+}$ occupy the $M$1 sites, forming the $Fe^{2+}$ chain along the $a$-axis.

The temperature dependence of magnetic susceptibilities of $FeTi_2O_5$ single crystal along different axes are displayed in Fig. 2. The rapid decrease at $T_N$ = 42 K in the $\chi(T)$ curve along the $b$-axis of $FeTi_2O_5$ corresponds to the long-range antiferromagnetic ordering of $Fe^{2+}$, consistent with the previous report [9], while the susceptibilities along the $a$-axis and the $c$-axis turn to increase with decreasing temperature below $T_N$. The anisotropic behavior of $\chi(T)$ curves suggests that the $b$-axis is the easy axis in $FeTi_2O_5$. When applying different magnetic fields along the $b$-axis of $FeTi_2O_5$, as shown in Fig. 2(b), $T_N$ decreases with increasing the field and the rapid drop in the magnetic susceptibility tends to disappear. The temperature dependence of magnetic susceptibility of $FeTi_2O_5$ above 90 K was fitted by the Pade approximation following Ref. [23]. The fitting yields nearest neighbor exchange interaction $J_{NN}$ ~ 22 K if the Heisenberg model is defined as

$$H_{Heis} = \sum_{i>j} J_{ij}\, S_i S_j \qquad (1)$$

and g-factors along different axes are $g_a$ ~ 2.24, $g_b$ ~ 2.20, and $g_c$ ~ 2.30, which is in accordance with the small difference in the susceptibilities for three different directions at 300 K. As shown in Fig.2 (c), by fitting the $1/\chi$-$T$ curve of $FeTi_2O_5$ with the Curie-Weiss law from 200 K to 300 K, the effective magnetic moments along the $a$, $b$, and $c$ axes were obtained as 5.96 $\mu_B$, 5.59 $\mu_B$, and 6.35 $\mu_B$, respectively, which are much larger than the spin-only value of 4.90 $\mu_B$ for the Fe$^{2+}$ with $S = 2$, indicating that there is an orbital contribution to the magnetic moment in FeTi$_2$O$_5$.

The field dependence of magnetization of FeTi$_2$O$_5$ presented in Fig. 2 (d) also confirmed the existence of the orbital contribution. Indeed, $M(H)$ curves along the $a$- and the $c$-axis of FeTi$_2$O$_5$ deviate from linearity at substantial fields and do not saturate even at about 100 kOe, suggesting a strong magnetic anisotropy due to unquenched orbital moment and substantial spin-orbit coupling. As mentioned above, the 1D chains of Fe$^{2+}$ will be formed along the $a$-axis in FeTi$_2$O$_5$ with highly ordered Fe$^{2+}$/Ti$^{4+}$ occupation. This quasi-one-dimensionality is manifested by the broad peak at around 75 K in the $\chi(T)$ curve, as shown in the inset of Fig. 2(a). Although the Fe$^{2+}$ chains exist in FeTi$_2$O$_5$, the magnetic structure of FeTi$_2$O$_5$ is 3D. This is also consistent with the high Neel temperature in FeTi$_2$O$_5$ ($T_N \sim 42$ K) which suggests a large interchain exchange interaction in FeTi$_2$O$_5$. However, care should be taken in this respect, since in some low-dimensional materials the ordering temperature is largely suppressed by the logarithm of ratio intra and interchain exchanges [24].

The large interchain exchange interaction in the FeTi$_2$O$_5$ having a triangular motif as shown in Fig. 1(a) can result in a geometrical frustration being antiferromagnetic. The frustration index commonly estimated as $|\theta_{CW}|/T_N$ varies from 2.9 to 5.0 for FeTi$_2$O$_5$ because the fitted Weiss temperature $\theta_{CW}$ differs along different axes. Lang *et al.* proposed that the long-range antiferromagnetic ordering in FeTi$_2$O$_5$ appears if the frustration is released by the spin Jahn-Teller effect [9], but no structural distortion has been detected so far.

## B. Thermal expansion, specific heat, and thermal conductivity

We measured the temperature dependence of thermal expansion ($\Delta L/L_{80K}$) of FeTi$_2$O$_5$ single crystal along different axes, where the value of $\Delta L/L_{80K}$ was normalized as $[L(T)-L(80K)]/L(80K)$, as shown in Figs. 3(a) and 3(b). As the temperature decreases, the $a$-axis of FeTi$_2$O$_5$ shows the positive thermal expansion behavior with two anomalies around $T_N$ at 41 K and 43 K, respectively, while the negative thermal expansion appears along the $b$-axis, and the slope of the $\Delta L/L_{80K}$ curve along the $b$-axis also changes around $T_N$. This anomaly is easier to see from the thermal expansion coefficient $\alpha(T) = d[\Delta L(T)/L_{80\,K}]/dT$ plot shown in the insets of Fig. 3 (a) and (b). We also performed the magnetostriction measurement to confirm the existence of the spin-lattice coupling in FeTi$_2$O$_5$. As shown in Figs. 3 (c) and (d), the negative and positive magnetostriction effects appear clearly below $T_N$ along the $a$- and $b$-axis, respectively, indicating that there is an obvious spin-lattice coupling in FeTi$_2$O$_5$. This magnetostriction can soften the phonons and induce lattice distortions in FeTi$_2$O$_5$ at $T_N$. A similar effect has been reported for some transition-metal oxides, e.g., for $Fe_3(PO_4)O_3$ [25], $Bi_2Fe_4O_9$ [21,26]. With decreasing the temperature, the magnitude of magnetostriction of $FeTi_2O_5$ increases, as shown in Fig.3(c), indicating the enhancement of spin-lattice coupling at low temperature.

The temperature dependence of specific heat of $FeTi_2O_5$ single crystal was shown in Fig. 4(a). A peak can be seen at $T_N$ = 42 K in the $C(T)$ curve, corresponding to the antiferromagnetic ordering of $FeTi_2O_5$ [9]. As the temperature decreases, a shoulder-type peak appears in the $C(T)$ curve at 41 K. These two peaks are consistent with two anomalies in thermal expansion along the $a$-axis. This double-peak behavior in $C(T)$ may correspond to two domains of comparable volume. It is tempting to ascribe this effect to two magnetic domains, characterized by propagation vectors $\mathbf{k_1} = \left(\frac{1}{2},\frac{1}{2},0\right)$ and $\mathbf{k_2} = \left(-\frac{1}{2},\frac{1}{2},0\right)$ [8]. These two phases are degenerate and tiny distortions can lift this degeneracy through the spin JT effect, resulting in the formation of several domains. However, one cannot rule out a simpler scenario, when domains appear due to different degrees of stoichiometry, which comes from the mutual substitution of $Fe^{2+}$ and $Ti^{4+}$. This slightly non-uniform distribution of $Fe^{2+}$ and $Ti^{4+}$ can cause local strains, stabilizing one or another magnetic phases. Finally, the peak at $T_N$ = 42 K can be associated with the long-ranged antiferromagnetic ordering, and the peak at 41 K comes from the structural distortion induced by the magnetostriction phenomenon, which is commonly seen in antiferromagnetic oxides below the Neel temperature, e.g. CoO *etc*. [27].

Magnetic contribution to the specific heat $C_m(T)$ of $FeTi_2O_5$ can be obtained by subtracting the lattice specific heat $C_p(T)$ from the total specific heat of $FeTi_2O_5$. The $C_p$ of the $FeTi_2O_5$ single crystal was estimated from the $C_p$ of the nonmagnetic $MgTi_2O_5$ single crystal by using the formula $C_p(T)_{FeTi_2O_5} = \sqrt{M_{FeTi_2O_5}/M_{MgTi_2O_5}}\, C_p(T)_{MgTi_2O_5}$, in which $M_{FeTi_2O_5}$ and $M_{MgTi_2O_5}$ represent the relative molecular masses of $FeTi_2O_5$ and $MgTi_2O_5$, respectively. The magnetic entropy change of $FeTi_2O_5$ calculated from $\Delta S = \int_{T_1}^{T_2} \frac{C_m}{T} dT$ is 6.41 J mol$^{-1}$ K$^{-1}$ at $T_N$, which is about 48% of $R\ln(2\times 2+1) \approx 13.37$ J mol$^{-1}$ K$^{-1}$ ($R$ is the gas constant) expected for $S=2$ and close to the value of $R\ln 2 \approx 5.76$. The difference between the calculated and the expected value of $\Delta S$ may come from two reasons: (i) the magnetic entropy is released far above $T_N$ by the short-range magnetic ordering in the 1D $Fe^{2+}$ spin chains and (ii) due to the spin JT effect.

As an exclusively sensitive probe for the itinerant excitations [28], the thermal conductivity of $FeTi_2O_5$ single crystal was also measured to further investigate the source of the structural distortions in $FeTi_2O_5$. In Fig. 5, the thermal conductivity of $FeTi_2O_5$ increases monotonically with decreasing temperature and suddenly enhances around $T_N$, which is more pronounced in the $1/\kappa(T)$ curve, and a cusp occurs at $T_N$. When applying the magnetic field of 120 kOe, the $\kappa$ of $FeTi_2O_5$ below $T_N$ was apparently suppressed. The change of the slope of $1/\kappa$ is commonly seen in the spinel oxides with orbital degrees of freedom and is explained as the result of the orbital ordering [29,30]. Therefore, one may expect that this change in the $1/\kappa(T)$ curve of $FeTi_2O_5$ also stems from the orbital ordering of $Fe^{2+}$, which leads to the anomaly in temperature-dependence of $\Delta L/L_{80K}$, but this hypothesis must be verified by other methods, e.g., X-ray spectroscopy, while our DFT calculations do not detect any changes in the orbital occupations, see next section. So, this slope change may imply the appearance of spin JT in $FeTi_2O_5$. Another possibility for the slope changing in $\kappa(T)$ is that the phonons are transferred more efficiently in the ordered phase. Since $FeTi_2O_5$ is a quasi-1D antiferromagnet with $S = 2$, a spin gap is expected to open in its magnetic excitation spectrum. Therefore, the increase of $\kappa$ below $T_N$ may also result from the weakening of the scattering of phonons by magnons. Similar effect has been observed in some quasi-1D antiferromagnets, e.g., $NaV_2O_5$ [31]. The low-temperature peak in the $\kappa$-$T$ curve of $FeTi_2O_5$ around 30 K is commonly seen in solids, resulting from the phonon-phonon scattering [32].

### C. First-principles calculations

To check possible modifications of the orbital structure in $FeTi_2O_5$ and study exchange interaction, we performed GGA and GGA+U calculations. It has to be mentioned that Fe resides in 4c positions in the *Cmcm* space group [9], which corresponds to the $C_{2v}$ point group. The degeneracy of *3d*-orbitals is completely lifted in this case. Indeed, results of the Wannier function projection of the non-magnetic GGA Hamiltonian using Wannier90 code [33] for the experimental crystal structure measured for room temperature [9] presented in Fig. 6a confirm general symmetry arguments. Thus, there is no orbital degeneracy to be lifted. The lowest in energy is the *xz* orbital (in the global coordinate system), which is shown in Fig. 6b. In the high-spin $d^6$ configuration of $Fe^{2+}$ a single electron with spin-minority must occupy this orbital, and the next in energy orbital is split by 210 meV. Such a large crystal-field splitting agrees with strong distortions of $FeO_6$ octahedra and disfavors any redistribution of electrons, which is needed for any non-trivial orbital ordering below $T_N$. Weak distortions due to the spin JT are not expected to strongly affect electronic level structure.

The direct GGA+U calculations confirm that correlation effects do not change the situation and indeed a single electron with spin-minority occupies the *xz* orbital, making this orbital doubly occupied and inactive in the exchange interaction along the chain. Moreover, we also performed optimization of the crystal structure and calculations taking into account experimental modifications of lattice parameters according to our thermal expansion measurements (expand along the *a*-axis and shrink in the *b*-direction). None of them show changes in the orbital structure.

Fig. 7 shows the density of state plots in the case of GGA+U calculation with spins in the chain ordered antiferromagnetically. One can see that the bottom of the conduction band is formed by completely empty Ti *3d* states, while the top of the valence band is by Fe *3d* spin-minority states (narrow bands just below the Fermi level have *xz* character; corresponding orbital is shown in Fig. 6b). The band gap is ~1.7 eV.

Finally, we calculated the exchange interaction parameter for the Heisenberg model (1) using method of 4 configurations [34]. The intra-chain exchange was found to be *J*=17.7 K. This agrees with the fitting of magnetic susceptibility by the Pade approximation of the 1D Heisenberg model as explained above. The diagonal inter-chain exchange (for simplicity we do not differentiate between such exchanges in the *ab* and *ac* planes) was calculated to be *J'*=0.6 K.

In addition, GGA+U+SOC (SOC stands for the spin-orbit coupling) calculations reveal that spins are directed along the *b*-axis with orbital momentum ~0.1$\mu_B$. This fully agrees with our magnetization measurements (see Sec. IIIa). The single-ion anisotropy modeled by $H_{SIA} = \sum_i A (S_i^z)^2$ was found to be *A=4 K* from the total energy GGA+U+SOC calculations of 4 configurations: when at one of the spins is directed along *b*, *-b*, *a*, or *-a*, while all other spins are along the *c* axis.

Thus, we see that FeTi$_2$O$_5$ exhibits the Ising-like behavior and can be an example of a quasi 1D Ising antiferromagnet with spin *S* = 2. Indeed, ratio of intra- to inter-chain exchanges *J/J'* ~ 30 and *A/J'* ~ 7. This model has highly unusual properties in the presence of the transverse magnetic field. There can be a quantum critical behavior (as the field increases) together with the exotic E$_8$ excitations. This is now one of the hot subjects in the physics of low-dimensional magnetic systems and material realizations of these phenomena e.g. in 1D ferromagnetic CoNb$_2$O$_6$ [35–37] or antiferromagnetic BaCo$_2$V$_2$O$_8$ [38–40] are under active studying. However, interest in the Ising model in the transverse field is not restricted by materials with *S*=1/2, since the field induces tunneling processes between states in a double-well potential characterized by different spin projections and these are not necessarily only the states with $S_z$=1/2 [41].

## IV. CONCLUSION

The pseudobrookite FeTi$_2$O$_5$ single crystal with highly ordered Fe$^{2+}$/Ti$^{4+}$ occupation was grown under the pure Ar atmosphere. Magnetostriction measurements clearly demonstrate the appearance of the magnetoelastic coupling, which is consistent with the previously proposed spin Jahn-Teller scenario of leaving degeneracy in this highly frustrated system due to coupling with crystal lattice and stabilizing a long-range magnetic order [9]. While tiny distortions associated with the spin Jahn-Teller effect are unluckily to be measured by diffraction techniques, our experiments reveal the existence of magnetoelastic coupling in $FeTi_2O_5$ and *ab initio* calculations rule out another mechanism of symmetry lowering due to orbital degrees of freedom. In addition, density function theory calculations unravel a substantial easy-axis anisotropy putting this material to the class of quasi 1D Ising model, which agrees with the field dependence of magnetization. Exotic properties of Ising systems in the transverse magnetic fields together with a strong magnetoelastic coupling make $FeTi_2O_5$ a unique object interesting for further detailed investigation.

## ACKNOWLEDGMENTS

We thank Valentin Irkhin, Alexander Vasiliev, Dylan Behr, and Roger Johnston for fruitful discussions and the Synergetic Extreme Condition User Facility (SECUF) for physical properties measurements. This work was supported by National Key R&D Program of China No. 2023YFA1406100, the National Natural Science Foundation of China (Grant No. 52372003) and the Funds from Beijing National Laboratory for Condensed Matter Physics. Work of S.V.S. was supported by "Quantum" program (№ 122021000038-7).

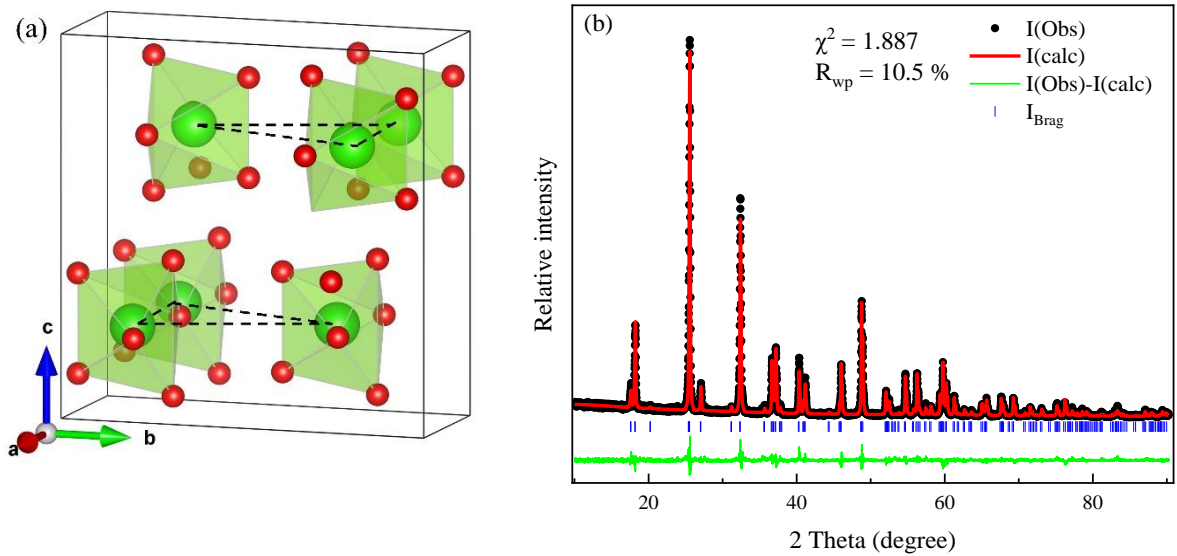

FIG.1 (a) The schematic diagram of the geometrical frustration in $FeTi_2O_5$, in which only $Fe^{2+}$ (the green ball) and the corner-shared $O^{2-}$ (the red ball) were shown. Fe chains run along *a*-axis. The dashed line shows the triangular geometric frustration in $FeTi_2O_5$. (b) The Rietveld refinement results of $FeTi_2O_5$.

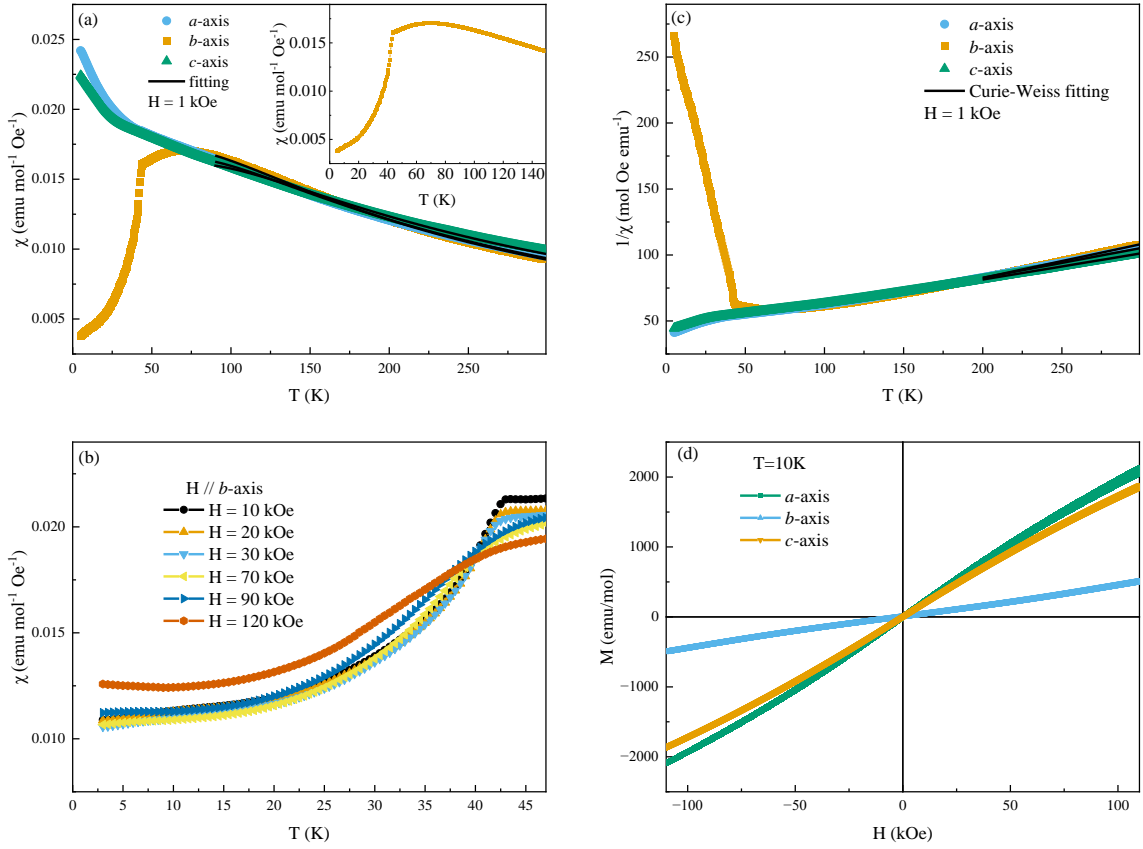

FIG.2 The magnetic measurements of $FeTi_2O_5$ single crystal along different axes. (a) $\chi$-$T$. The inset presents the broad peak along the $b$-axis in enlarged form. (b) Low-temperature susceptibilities at different fields along the $b$-axis. (c) $1/\chi$-$T$. (d) $M(H)$ curves.

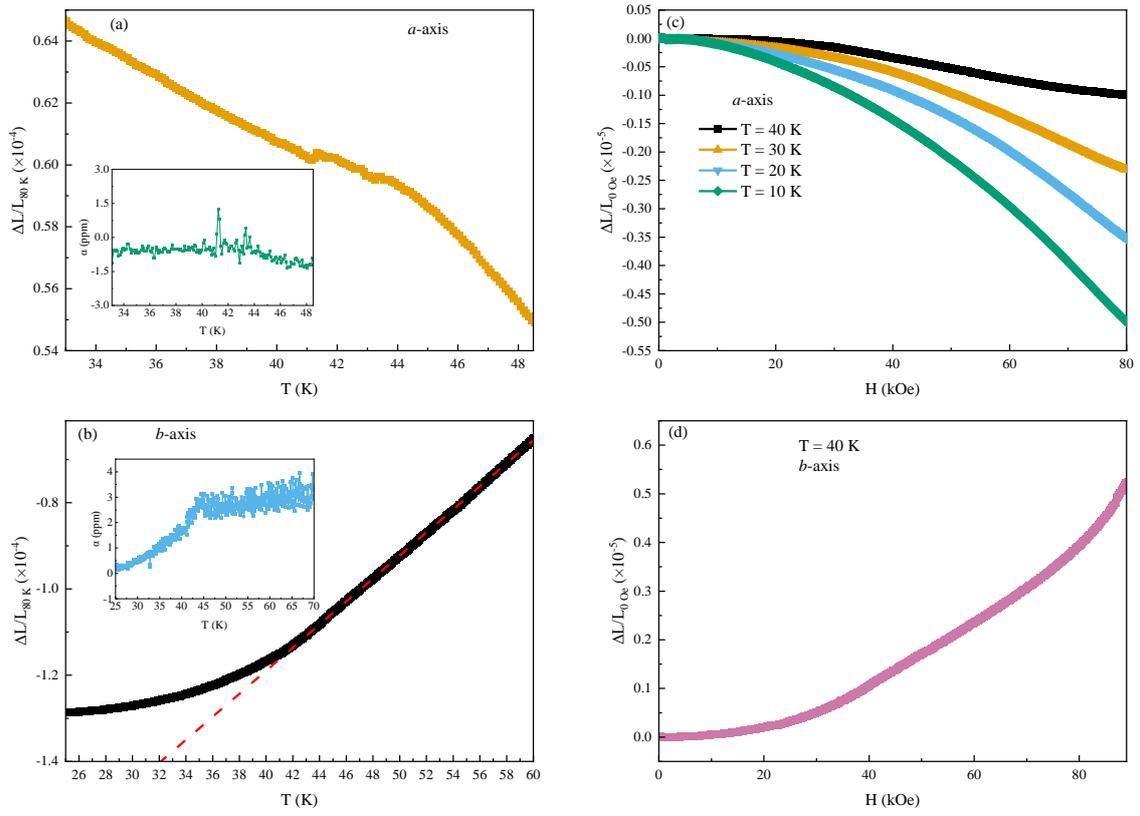

FIG.3 (a) and (b) The thermal expansion of FeTi$_2$O$_5$ single crystal along the *a*-axis and *b*-axis. The insets show the thermal expansion coefficient *α* of FeTi$_2$O$_5$. (c) The magnetostriction of FeTi$_2$O$_5$ single crystal along the *a*-axis at different temperatures. (d) The magnetostriction of FeTi$_2$O$_5$ single crystal along the *b*-axis below $T_N$.

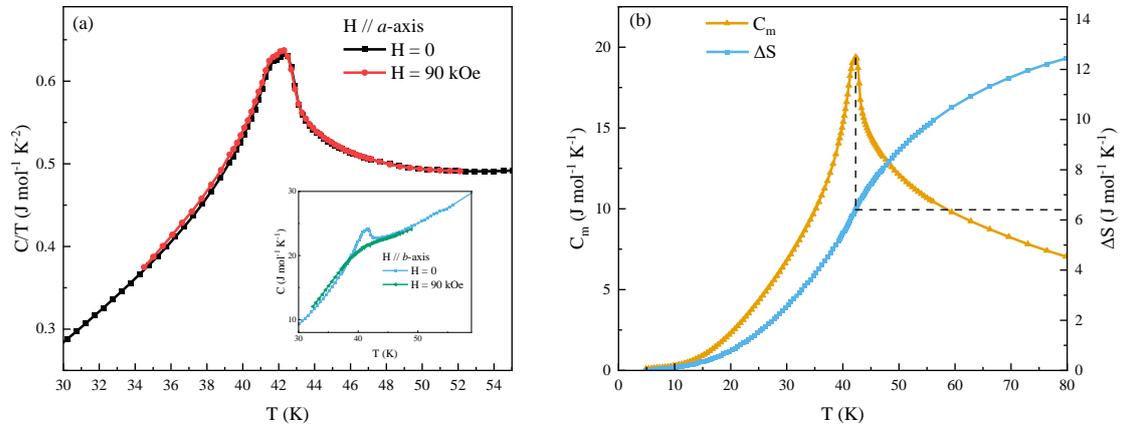

FIG. 4 (a) The specific heat of FeTi$_2$O$_5$ under different magnetic fields. (b) The magnetic specific heat and change of magnetic entropy of FeTi$_2$O$_5$.

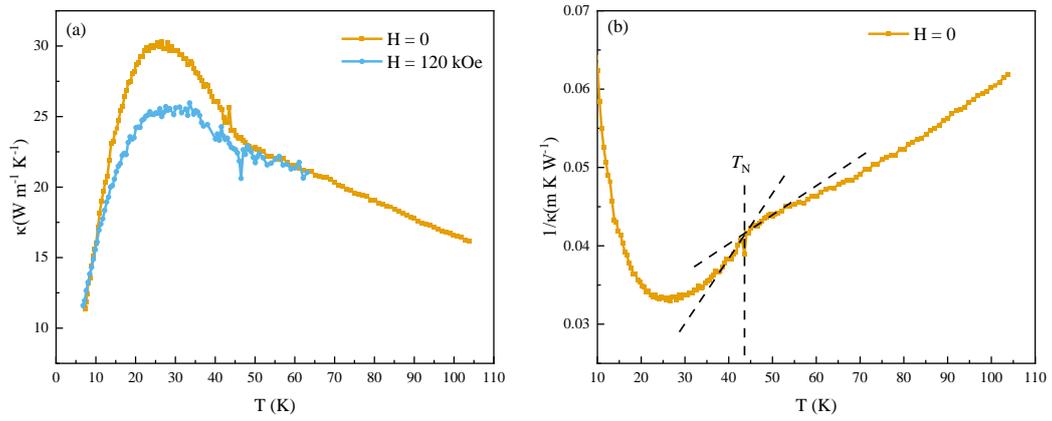

FIG. 5 The thermal conductivity of FeTi$_2$O$_5$ single crystal along the *a*-axis under different magnetic fields. (a) *κ-T*, (b) 1/*κ-T*. The dashed line in the 1/*κ-T* at zero field is used to show the change in the slope.

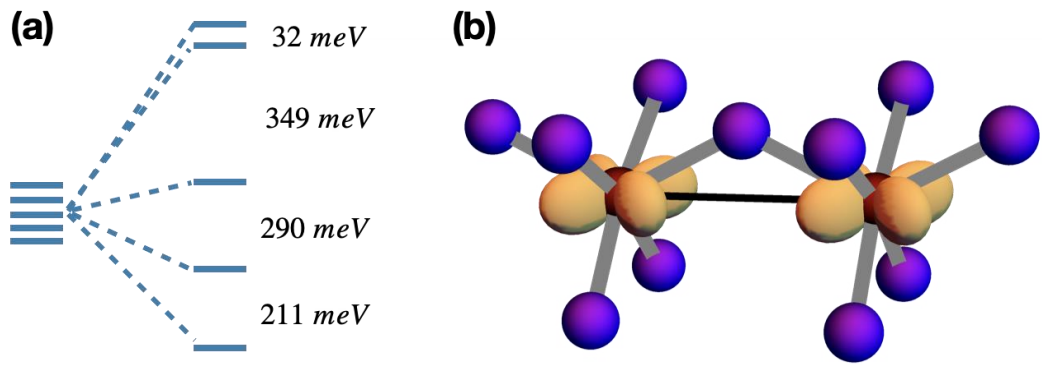

Fig. 6 (a) Crystal-field splitting of the Fe *3d* shell as obtained by the Wannier function projection of non-magnetic GGA calculations. (b) the lowest in energy *xz* orbital (in the global coordinate system) in GGA non-magnetic calculations; Fe ions are shown by red, while O is by violet balls.

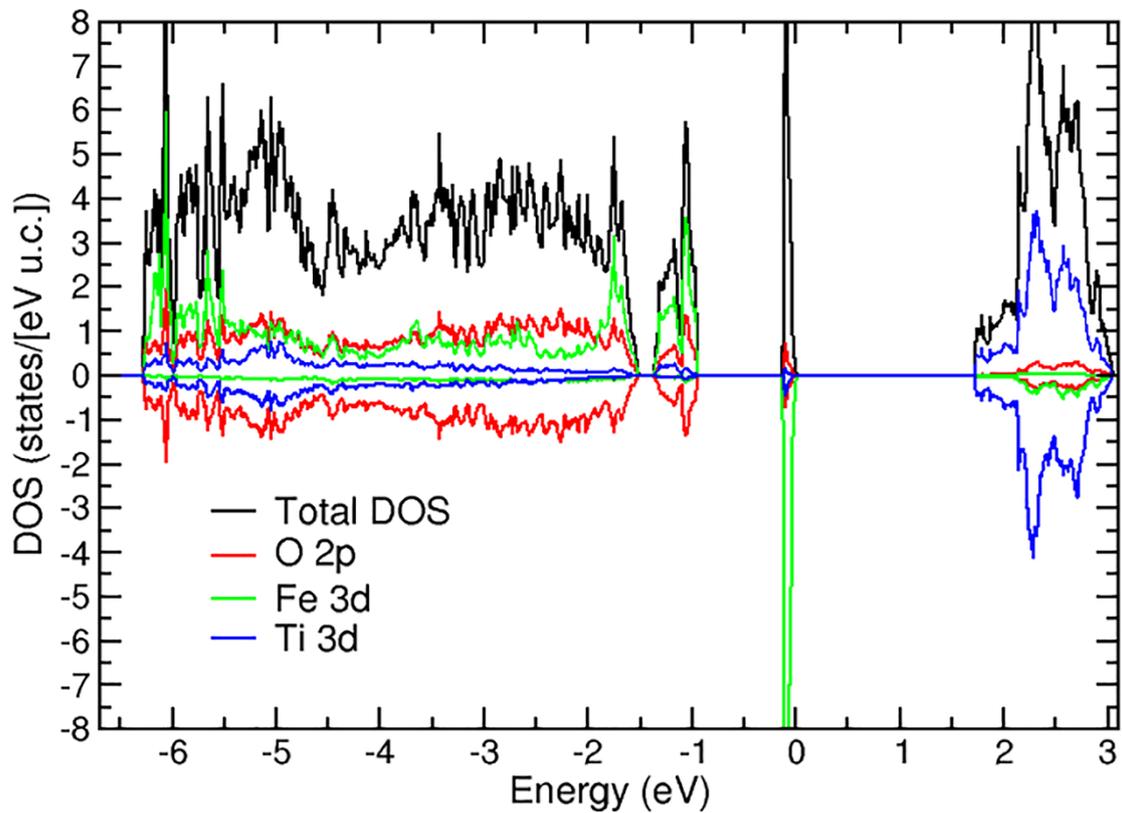

Fig. 7 Total and partial densities of states (DOS) as obtained in GGA+U calculations for the magnetic configuration, where spins in the same chain are AFM ordered. For O *2p*, Fe *3d*, and Ti *3d* two spin projections are shown by (positive DOS for spin majority, negative for spin minority). The Fermi level is set to zero.

TABLE I. Refined crystal structure parameters of FeTi$_2$O$_5$

Cell parameters
Space group: *Cmcm*
*a,b,c* (Å) 3.74014(2) 9.76375(6) 10.08496(7)
Atomic fractional coordinates

| Atom | Site | a | b | c | Occ. |
| --- | --- | --- | --- | --- | --- |
| Fe1 | 4c | 0 | 0.1916 | 0.2500 | 0.8885 |
| Ti1 | 4c | 0 | 0.1916 | 0.2500 | 0.1115 |
| Fe2 | 8f | 0 | 0.1343 | 0.5666 | 0.0558 |
| Ti2 | 8f | 0 | 0.1343 | 0.5666 | 0.9442 |
| O1 | 4c | 0 | 0.7810 | 0.2500 | 1.0000 |
| O2 | 8f | 0 | 0.0450 | 0.1130 | 1.0000 |
| O3 | 8f | 0 | 0.3150 | 0.0610 | 1.0000 |